\documentclass[12pt,preprint]{aastex}

\usepackage{emulateapj5}

\newcommand{\kms}{km~s$^{-1}$}
\newcommand{\lya}{Lyman~$\alpha$}
\newcommand{\lyb}{Lyman~$\beta$}
\newcommand{\lyg}{Lyman~$\gamma$}
\newcommand{\lyd}{Lyman~$\delta$}
\newcommand{\lye}{Lyman~$\epsilon$}

\newcommand{\chid}{$\chi^2$}

\input{psfig}

\begin{document}


\title{Hydrogen column density evaluations toward Capella:\\ 
consequences on the interstellar deuterium abundance}



\author{A.~Vidal--Madjar, R.~Ferlet}
\affil{Institut d'Astrophysique de Paris, CNRS, 98 bis bld Arago, 
F-75014 Paris, France\\ vidalmadjar@iap.fr~; ferlet@iap.fr}


%\altaffiltext{1}{Visiting Astronomer, Cerro Tololo Inter-American 
%Observatory. CTIO is operated by AURA, Inc.\ under contract to 
%the National Science Foundation.}
%\altaffiltext{2}{Society of Fellows, Harvard University.}
%\altaffiltext{3}{present address: Center for Astrophysics,
%    60 Garden Street, Cambridge, MA 02138}

%\altaffiltext{4}{Visiting Programmer, Space Telescope Science Institute}
%\altaffiltext{5}{Patron, Alonso's Bar and Grill}


\begin{abstract}

The deuterium abundance evaluation in the direction of Capella has for a long
time been used as a reference for the local interstellar medium (ISM) within
our Galaxy. We show here that broad and weak H$\,${\sc i} components could be 
present on the Capella line of sight, leading to a large new additional
systematic uncertainty on the $N$(H$\,${\sc i}) evaluation.

The D/H ratio toward Capella is found to be equal to
$1.67~(\pm0.3)\times10^{-5}$
with almost identical \chid\ for all 
the fits (this range includes only the systematic error~; the 2
$\sigma$~statistical one is almost negligible in comparison). It is concluded 
that D/H evaluations over H$\,${\sc i} column densities below $10^{19}~cm^{-2}$
(even perhaps below $10^{20}~cm^{-2}$ if demonstrated by additional 
observations) may present larger uncertainties than previously anticipated. 
It is mentionned that the D/O ratio might be a better tracer for D$\,${\sc i} 
variations in the ISM as recently measured by the Far Ultraviolet 
Spectroscopic Explorer ({\it FUSE}).

\end{abstract}




\section{Introduction}

Deuterium is understood to be only produced in significant amount during 
primordial Big Bang nucleosynthesis (BBN) and thoroughly destroyed in stellar 
interiors. Deuterium is thus a key element in cosmology and in galactic 
chemical evolution (see e.g. Audouze \& Tinsley 1976). Indeed, its primordial 
abundance is the best tracer of the baryonic density parameter of the Universe
$\Omega_B$, and the decrease of its abundance during the galactic evolution 
should trace the amount of star formation (among other astrophysical
interests).

In the Galactic ISM, D/H measurements made toward hot stars have suggested 
variations~: IMAPS observations toward $\delta$~Ori led to a low value 
(Jenkins {\it et al.}~1999), confirming the previous analysis by Laurent 
{\it et al.} (1979) from {\it Copernicus} observations, while toward 
$\gamma^2$~Vel they led to a high value (Sonneborn {\it et al.}~2000).
This seems to indicate that in the ISM, within few hundred parsecs, D/H may 
vary by more than a factor $\simeq3$.

In the nearby ISM, the case of G191--B2B was studied in detail (see the most 
recent analysis by Lemoine {\it et~al.} 2002) and the evaluation toward 
Capella (Linsky {\it et~al.} 1995) taken as a reference. Their comparison 
provided, for a while, a possible case for D/H variations within the local ISM.

Concerning G191--B2B, Lemoine {\it et~al.} (2002) have shown that the total 
$N_{\rm tot}$(H$\,${\sc i}) column density evaluation was greatly perturbed 
by the possible addition of two broad and weak H$\,${\sc i} components. Such 
components, able to mimic the shape of the \lya\ damping wings, can induce an 
important decrease of the evaluated 
$N_{\rm tot}$(H$\,${\sc i}). To illustrate this point, 
the error bar estimation on $N_{\rm tot}$(H$\,${\sc i}) from all previously
published studies considered as the extremes of a 2$\sigma$~limit was of the 
order of dex~0.07, while including the Lemoine {\it et~al.} (2002) analysis 
enlarged the error bar to about dex~0.37. This huge change has, of course, a 
considerable impact on any D/H evaluation.

This raises two crucial questions. First, is that situation typical of 
G191--B2B alone and possibly due to an unexpected shape of the core of the 
stellar \lya\ profile improperly described by the theoretical models? Second, 
if weak H$\,${\sc i} features are present in the ISM, to what extent are 
evaluations toward other targets affected?

\section{Summary of the G191--B2B case}

From the combination of {\it STIS} echelle observations (spectrograph on board 
the Hubble Space Telescope, HST) and {\it FUSE} ones (the Far Ultraviolet
Spectroscopic Explorer, Moos {\it et al.}, 2000), Lemoine {\it et al.} (2002) 
have found through iterative fitting process (with the {\tt Owens.f} fitting 
program developed by Martin Lemoine and the French FUSE team) that three 
interstellar absorption components are present along the line of sight and that
two additional broad and weak H$\,${\sc i} components could be added, detected 
only over the Lyman~$\alpha$ line (negligible over the \lyb\ line) but 
important enough to strongly perturb the total H$\,${\sc i} column density 
evaluation. 

Within the local ISM, it has been shown that such additional H{\sc i} 
absorptions are often present~; they have been interpreted either as cloud 
interfaces with the hot gas within the local ISM (Bertin {\it et al} 1995) or 
as ``hydrogen walls'', signature of the shock interaction between the solar 
wind (or stellar wind) and the surrounding ISM (Linsky, 1998). This latter 
heliospheric absorption has been modeled by Wood {\it et al.} (2000) and a 
prediction derived in the direction of G191--B2B (see 
Figure 9 of Lemoine {\it et al.} 
2002). Most of the predicted absorption is expected in the saturated core of 
the observed interstellar line but some weak absorption ($\sim5\%$~of the 
continuum) might extend over several tenths of angstroms on the red side of 
the line, due to the neutral hydrogen atoms seen behind the shock in the 
downwind direction where G191--B2B is located. It was found that the 
combination of two broad and weak H{\sc i} components can easily reproduce the 
model prediction. If real, besides the three interstellar absorptions, a 
fourth component representing the bulk of the predicted absorption and a fifth 
one for the broad and shallow extended red wing are needed. This is exactly 
what Lemoine {\it et al.} (2002) have found.

In the course of determining the minimum number of components (each defined by 
its H{\sc i} column density $N$, its velocity $v$, its temperature $T$ and 
turbulence broadening $\xi$) needed to fit the data, Lemoine {\it et al.} 
(2002) completed the $F-$test which uses the Fisher-Snedecor law describing
the probability distribution of $\chi^2$ ratio. What is tested is the
probability that the decrease of the $\chi^2$ with additional components 
is not simply due to the increase of free parameters. The result gives a 
probability $\leq10^{-4}$ and  $\sim5~10^{-3}$ that a fourth and a fifth 
H{\sc i} component are respectively not required by the data. These low
probabilities of non occurence strongly suggest that Lemoine {\it et al.} 
(2002) have indeed detected the heliospheric absorption downwind in the 
direction of G191--B2B.

Note however that this heliospheric complex absorption profile is simulated by 
two components whose physical meaning in terms of hydrogen content and/or 
temperature is not clear. Furthermore, the photospheric \lya\ stellar core is 
difficult to evaluate (see discussion in e.g. Lemoine {\it et al.} 2002) and 
is slightly red-shifted relative to the ISM absorptions~; this result may very 
well be simply related to the use of a white dwarf as background target star. 
The detailed analysis of the Capella line of sight could directly test the
heliospheric hypothesis.

\section{The case of Capella}

If the two additional components present along the G191--B2B line of sight are 
as a matter of fact due to an heliospheric phenomenon, it is an extremely local 
signature (within few hundreds of astronomical units to be compared to the few 
tens of parsecs lines of sight lengths) which should be also present along the 
Capella sight-line, both stars being separated by only $7^\circ$ on the sky,
and similar in shape to the structure predicted and observed in the direction 
of G191--B2B. If that description is correct, we are expecting an extra 
absorption reasonably represented by two additional components, a main one 
mostly lost within the ISM absorption core and a weak one extending over 
several tenths of angstroms on the red side of the line, again due to the 
neutral hydrogen atoms seen behind the shock in the downwind direction where 
both G191--B2B and Capella are located.

Recently, Young {\it et al.} (2002) analysed new obervations obtained at \lyb,
\lyg\ and the whole Lyman series with {\it FUSE}. The precise \lya\ stellar 
profile compatible with all Lyman lines and with the data sets obtained at 
different phases of the Capella binary system (see also Linsky {\it et al.} 
1995) was reevaluated (Wood, 2001) and is used here as a reference profile, 
S$_{\rm R}$.

\begin{figure*}
\begin{center}
\psfig{figure=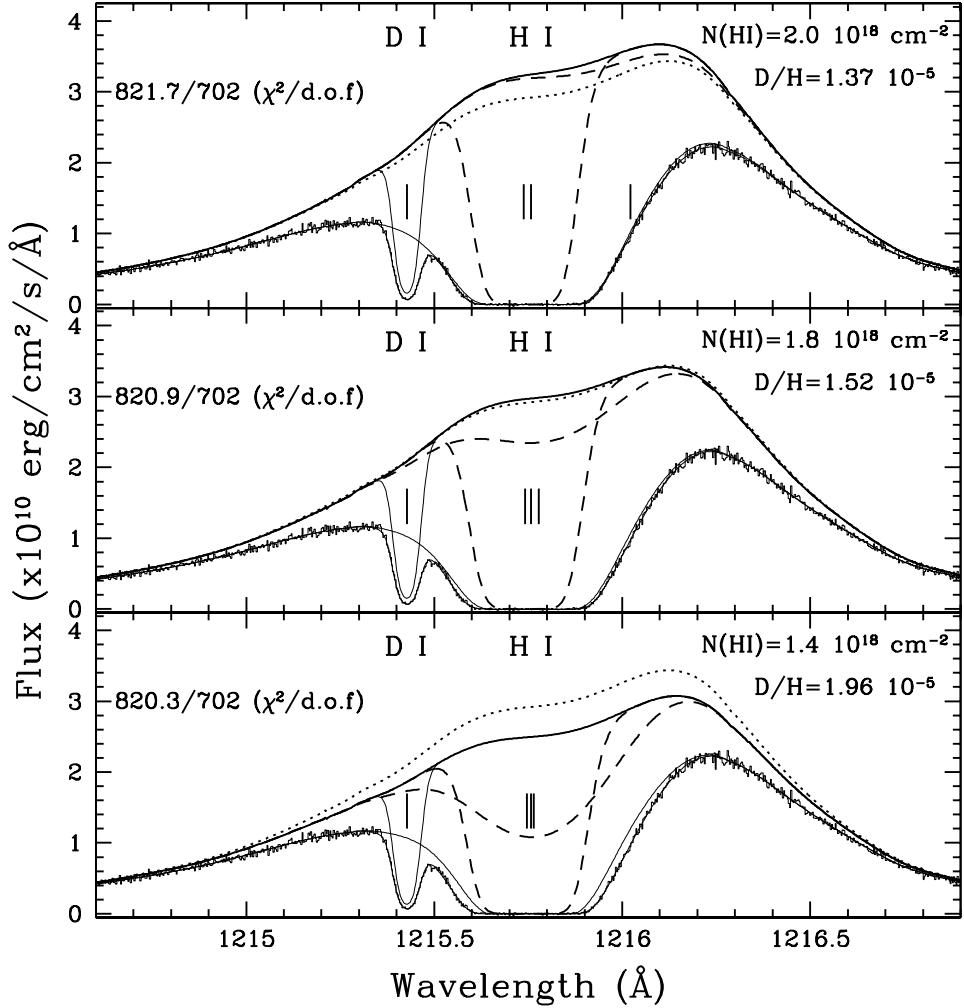,width=14.5cm,angle=0}
\caption[]{ {\it GHRS} observations of the Capella \lya\ line (thin histogram).
Since Capella is close to G191--B2B on the sky, two broad and weak H{\sc i} 
absorption components (thick dashed lines) were freely added in the fitting 
process. The outcome is very similar to the one found for the G191--B2B line
of sight. In each pannel, from top to bottom, the H{\sc i} column density of 
the unique ISM component is fixed at different values (noted on the upper
right corners). The fits are shown as
a thin almost invisible line running through the data points (according to
the very high quality of these fits) in all three pannels. 
Since these fits are very
similar in terms of \chid\ it is impossible to differentiate one from the
other by eye. What is different is shown as thin lines corresponding to the
individual contributions of the H{\sc i} and D{\sc i} ISM components. 
Vertical tick marks 
give the spectral positions of each component. The reference stellar continuum 
S$_{\rm R}$ (thick dotted line) as evaluated by Wood (2001) from the 
adjustment of several data sets is shown along with the stellar continuum 
(S$_{\rm R}$ corrected by a 8$^{\rm th}$ order polynomial, thick solid line) 
found 
in the fitting process. If the stellar profile is not known to better than 
$\pm 10\%$, the H{\sc i} column density on the line of sight is poorly 
evaluated as well as the D/H ratio.}
\label{Cap-contVar}
\end{center}
\end{figure*}

We thus revisited the fits completed over the \lya\ line as observed toward 
Capella with the best available data set, i.e. the one obtained with the
{\it GHRS} (the Goddard High Resolution Spectrograph on board {\it HST}). The 
study by Linsky {\it et al.} (1995), essentially confirmed by Vidal--Madjar 
{\it et al.} (1998), shows that only one interstellar component (the Local 
Interstellar Cloud, LIC, also seen toward G191--B2B) is needed on that line of 
sight. This very simple structure strengthens the Capella case as the simplest 
one where D/H can be very well evaluated. However, Vidal--Madjar {\it et al.} 
(1998) have already noted that an additional weak and broad H{\sc i} component 
was required to better reproduce the profile~; this was a first indication of 
the presence of an heliospheric absorption toward Capella. In fact, we were 
able to show that, as in the case of G191--B2B, the addition of one or two 
weak and broad H{\sc i} components (together with the very weak geocoronal 
component present at a known velocity but not shown on Figure~\ref{Cap-contVar}
for clarity) improves the \chid. More precisely we fitted the GHRS data 
assuming that the stellar continuum was S$_{\rm R}$. Adding successively to 
the fit one then two free H{\sc i} components (the added H{\sc i} components 
have only three free parameters, velocity $v$, column density $N$ and width 
{\bf T}, since the thermal $T$ or turbulent broadening $\xi$ act in an 
undifferentiated manner when only one species is observed) we obtained the 
following \chid/degree of freedom(d.o.f.) values~: for only the LIC component 
and the geocorona, 844.89/716~; for one additional component, 831.17/713~; for 
two additional components, 822.68/710. The $F-$test probabilities that these 
two additional components are not required by the data are respectively 
$8.5~10^{-3}$ and $6.3~10^{-2}$. The first one is clearly needed here (its 
correlated parameter ranges according to different possible solutions similar 
in terms of \chid\ are~:
$12\le$~$v$~(\kms)~$\le22$~; 
$1.4~10^{14}\le$~$N$~(cm$^{-2}$)~$\le4.4~10^{15}$~; 
$10000\le$~{\bf T}~(K)~$\le32000$)
but unlike in the case of G191--B2B, the second one corresponding to the 
weaker and broader one (parameter ranges are~: 
$24\le$~$v$~(\kms)~$\le40$~; 
$3.6~10^{12}\le$~$N$~(cm$^{-2}$)~$\le8.5~10^{13}$~; 
$120000\le$~{\bf T}~(K)~$\le300000$) is less strongly needed. These ranges
are certainly compatible with the corresponding estimated values in the
direction of G191--B2B (see Figure 10 of Lemoine {\it et al.} 2002).

To search for the possible impact of the choice of the continuum on the 
evaluation of $N$(H{\sc i}), we fixed this value and looked for the best 
fitted solutions while the stellar continuum we used S$_{\rm R}$~was allowed 
for some variations by multiplying it by a low order polynomial 
(8$^{\rm th}$~order) which coefficients were free to vary along with all
components parameters. Results are shown in Figure~\ref{Cap-contVar}. Slight 
changes of the continuum shape by no more than $\pm 10\%$~lead to nearly 
identical \chid\ values, with $N$(H{\sc i}) varying from 
$2.0\times10^{18}~cm^{-2}$ to $1.4\times10^{18}~cm^{-2}$ which corresponds to 
a change in D/H from $1.37\times10^{-5}$ to $1.96\times10^{-5}$. This is 
clearly a larger range ($\sim \pm 0.3\times10^{-5}$) than the one previously 
claimed ($\le \pm 0.2\times10^{-5}$, Linsky {\it et al.} 1995). 

The situation could be even worse since we do not know how far the Capella 
continuum could be away from S$_{\rm R}$. The question is thus to evaluate if 
the Capella \lya\ stellar continuum shape is estimated to better than 
$\pm 10\%$~or not~? It is true that having a binary system can help 
constraining the continuum shape as Linsky {\it et al.} (1995) did, but their 
whole approach requires that the \lya\ stellar profiles of both G1~III and 
G8~III stars are invariant with phase and time. In fact, from the study of 120 
{\it IUE} echelle spectra, Ayres {\it et al.} (1993) have shown on one hand 
that the line fluxes were surprisingly stable, but on the other hand that 
whichever way they process the data, obvious variations were seen. These seem 
to be related to variations of the blue peak of the G1~III dominant stellar 
\lya\ line. They found that in the 1981--1986 interval, the line shape at 
phase 0.25 of the system was quite stable and similar to the one recorded with 
the {\it GHRS} in 1991 (at a $\pm 10\%$ level). Earlier spectra taken in 1980 
or later ones observed after 1986 look quite different. This very careful 
study shows that with the {\it IUE} sensitivity level of $\pm 5\%$, variations 
are clearly detected. Since Linsky {\it et al.} (1995) used {\it GHRS} 
observations at two different phases of the system (0.26 and 0.80) taken 
respectively in April 1991 and in September 1993, i.e. two and a half year 
apart, it is difficult to ascertain that the \lya\ profile evaluated for each 
stellar component is well controlled. Because of the very careful analysis 
made by Linsky {\it et al.} (1995), it may be possible that the stellar \lya\ 
profiles are relatively well evaluated but certainly not at a level better 
than $\pm 10\%$~as previously mentioned.

Thus, an heliospheric absorption is also detected on the Capella sight--line~; 
furthermore, even in such a simple ISM configuration (a unique component), it 
appears impossible to tightly constrain the total H{\sc i} column density in 
that direction.

\section{Discussion}

We have shown that, for two lines of sight, $N$(H{\sc i}) cannot be evaluated 
with a high accuracy. Column densities on both sight--lines are very similar,
of the order of few times $10^{18}~cm^{-2}$. For lower column densities, the 
situation should be worse since then the possible absorption signature of the 
weak components is becoming relatively more and more important and the \lya\ 
line is getting closer to the flat part of the curve of growth where column
densities are indeed difficult to evaluate. 

Note however that the HST/EUVE comparison completed by Linsky {\it et al.}
(2000) shows that often $N$(H{\sc i}) values derived from the GHRS and EUVE 
data (not sensitive to weak H{\sc i}) are in good agreement, implying that 
heliospheric absorption (or other hot components) don't necessarily ruin 
Lyman-alpha analyses in a dramatic way. But clearly counter examples leave 
that question open due to possible systematics related to the evaluation of 
total H{\sc i} below the Lyman limit in the EUVE domain.

On the contrary, one could guess that for larger column densities the situation
should improve since the \lya\ damping wings are becoming broader and the 
signature of the weak features may disapear in the line core. Just above 
$10^{19}~cm^{-2}$, the reliability of the D/H values is greatly enhanced if 
the studied gas is demonstrably warm (6000 K~; for a thorough discussion see 
York, 2001)~; as one goes above $10^{20}~cm^{-2}$, credibility increases unless 
either cold gas components are hidden in the warm D{\sc i} but still affect 
the H{\sc i} damping wings or weak H{\sc i} features at high velocity are 
present.
 
The unknown referee further stressed this point through an impressive report.
He mentionned that 
at $N$(H{\sc i})=$2.\times10^{19}~cm^{-2}$, the half-intensity point of pure 
damping \lya\ and \lyb\ profiles, located respectively at velocity shifts of 
274~\kms\ and 55~\kms , should be the place where a putative high-velocity 
feature could have a strong perturbing influence on the damping profile. 
However only very few high velocity ISM components were detected above 120~\kms. 
This could lead to the inverse impression that for larger column densities, 
the estimation through the \lyb\ line should be more questionable than the one
made at \lya. As a matter of fact, in both the $\gamma$~Cas and $\zeta$~Pup
lines of sight, the H{\sc i} column density was discrepant when
derived through the \lya\ or the \lyb\ line (see below). 
However in both cases the \lyb\ 
estimations of $N$(H{\sc i}) are smaller, in contradiction with the formerly
suggested cause since the most perturbed evaluation by additional absorptions
should lead instead to larger column densities. 

High velocity ISM components essentially 
observed below 120~\kms were only searched for
through other lines and species than H{\sc i} at \lya\, the strongest
transition of the most abundant element. For instance, Cowie {\it et al.} 
(1979)~reported \lyd\ and \lye\ H{\sc i}
ISM absorptions up to about 105 km/s for 
$\iota$~Ori. From their study, the referee evaluated that a shock at 274~\kms\ 
should produce either a very broad ($\sim 175$~\kms) and undetectable (maximum 
depth of $5.9\times10^{-3}$) post-shock absorption signature or should 
originate in a region far downstream from the front where the gas has cooled 
and compressed enough to allow recombination of the H atoms, i.e. a shock from
a supernova explosion entering the radiative phase. In this second case 
however, he estimated from Cowie and York (1978) and Spitzer (1978) that such
signatures should occur only very near a known supernova event, i.e. within
about 30 to 60 pc for standard ISM and SN values. Thus, that looks unlikely 
too.

On the other hand, high-velocity gas could be generated by the target stars.
While Gry, Lamers and Vidal--Madjar (1984) seem to detect most of the activity
at velocities below 100 km/s, they nevertheless identified a transient 
component at -150 km/s toward $\gamma^2$~Vel through \lyd\ and another one at 
-220 km/s toward $\iota$~Ori, through \lye . Note also that this survey was
completed over a limited spectral domain scanned with the Copernicus
instrument, and not at \lya , {\it i.e.} with a relatively limited sensitivity.

One might argue that there are some risks that stellar ejecta could influence 
the \lya\ measurements~; but the observers have a good defense~: multiple
observations at very different epochs. This strategy was invoked by Jenkins 
{\it et al.} (1999) and Sonneborn {\it et al.} (2000) in their studies of
D/H toward $\delta$~Ori, $\zeta$~Pup and $\gamma^2$~Vel. Their findings are
thus pretty convincing in this regard.

All the above stated arguments should mitigate our concern that small amounts 
of H{\sc i} at high velocities are a likely source of confusion for the flanks 
of the damping profiles for $N$(H{\sc i}) of the order of or greater than 
$2.\times10^{19}$~cm$^{-2}$. 

One however should recall the two lines of sight for which the H{\sc i} column 
density was discrepant when derived through the \lya\ or the \lyb\ line~:

\begin{itemize}
\item
$\gamma$~Cas
\par 

  in Bohlin, Savage and Drake (1978), \lya\ only ({\it Copernicus})                                             
$N$(H{\sc i})~=~$1.45~(\pm0.29)\times10^{20}~cm^{-2}$
\par 
  in Ferlet {\it et al.} (1980), core of \lyb\ only ({\it Copernicus})
$N$(H{\sc i})~=~$1.10~(\pm0.10)\times10^{20}~cm^{-2}$
\par 
  in Diplas and Savage (1994), from \lya\ only ({\it IUE})
$N$(H{\sc i})~=~$1.48~(\pm0.31)\times10^{20}~cm^{-2}$
\par 

\item
$\zeta$~Pup
\par 

  in Bohlin (1975) from \lya\ only ({\it Copernicus})
$N$(H{\sc i})~=~$0.97~(\pm0.05)\times10^{20}~cm^{-2}$
\par 
  in Vidal--Madjar {\it et al.} (1977), from \lyb\ only ({\it Copernicus})
$N$(H{\sc i})~=~$0.80~(^{+0.1}_{-0.2})\times10^{20}~cm^{-2}$
\par 
  in Diplas and Savage (1994), from \lya\ only ({\it IUE})
$N$(H{\sc i})~=~$0.89~(\pm0.17)\times10^{20}~cm^{-2}$
\par 
  in Sonneborn {\it et al.} (2000), from \lya\ only ({\it IUE})
$N$(H{\sc i})~=~$0.92~(\pm0.01)\times10^{20}~cm^{-2}$
\par 
\end{itemize}
\par 

Since \lyb\ is less sensitive to weak H{\sc i} features, it is interesting to 
note that in both cases the \lyb\ evaluation is slightly below the \lya\ one, 
a possible indication of a similar effect for column densities of the order of 
$10^{20}~cm^{-2}$. Note however that these two cases are marginally convincing 
since the different evaluations are still compatible within the error bars.

Therefore, if this effect is indeed real for relatively large column
densities, it means that some of the D/H values 
may be underestimated and thus the higher D/H ratios may be favoured. This 
further shows the importance of the $\gamma ^2$~Vel estimation (Sonneborn 
{\it et al.} 2000).
 
\section{Conclusion}

According to our conclusion that there is lower precision for $N$(H{\sc i})
measurements within the local ISM, i.e. at low column densities, which induces 
large error bars on the corresponding D/H estimations:\\

\begin{itemize}
\item
the $N$(H{\sc i}) evaluation in the direction of Capella is relatively 
less accurate
than previously claimed and is of the order of 
log~$N$(H{\sc i})~$=18.22~(\pm0.08)$
leading to D/H~$=1.67~(\pm0.3)\times10^{-5}$~;
\item
an average D/H ratio may exist in the local ISM but should be larger than 
previously evaluated since locally $N$(H{\sc i}) could be overestimated by as 
much as about 20\%~; this does affect arguments about local variability~;
\item
D/O should be a better tracer of D variations as originally suggested by 
Timmes {\it et al.} (1997), and directly verified in the LISM  
(Moos {\it et al.} 2002~; H\'ebrard {\it et al.} 2002a, 2002b) and 
further confirmed by the stability of O/H over longer path length (Meyer 
{\it et al.} 1998~; Andr\'e {\it et al.} 2002), provided all sources of 
errors related to the oxygen measurements are well understood. 
\end{itemize}
\par 

Finally, we note also that a similar systematic effect has been pointed out by
Pettini and Bowen (2001) for the evaluation of D/H in quasars absorption line
systems. Again as in our study, the systems presenting the highest values of 
$N$(H{\sc i}) are derived from the damping wings of the \lya\ line, which also
includes all H{\sc i} in close proximity of the hydrogen at the line center of 
deuterium (Burles, 2001), while those for which $N$(H{\sc i}) is evaluated from 
the discontinuity at the Lyman limit present smaller column densities and 
larger D/H evaluations.

Clearly, many more lines of sights should be analysed to resolve this issue.

%-------------------- acknowledgments -----------------
\acknowledgments
%-----------------------------------------------------
{\bf Acknowledgments.} 

We would like to thank Brian Wood who kindly provided to us the Capella \lya\ 
stellar profile he has evaluated for his own study of that line of sight as
well as Jeff Linsky and Jeff Kruk for constructive comments. We 
are also pleased to warmly thank Don York for about twenty seven years of 
exciting collaboration and Martin Lemoine for the last decade of common 
enlightening work. We deeply thank our unknown referee whose report
was nearly as long as this letter and briefly summarized here.

%%%%%%%%%%%%%%%%%%%%%%%%%%%%%%%%%%%%%%%%%%%%%%%%%%%%%%%%%%%%%%%% 
%

\clearpage

\end{document}